\documentclass[reprint,aps,pra,amsmath,amssymb,showpacs]{revtex4-1}
\usepackage{graphicx}
\usepackage{dcolumn}
\usepackage{bm}

\begin{document}

\title{Photoionization of atoms and molecules studied by Crank-Nicolson method}

\author{Xue-Bin Bian}
\email{xuebin.bian@wipm.ac.cn}
\affiliation{State Key Laboratory of Magnetic Resonance and Atomic and Molecular Physics, Wuhan Institute of Physics and Mathematics, Chinese Academy of Sciences, Wuhan 430071, People's Republic of China}
\date{\today}

\begin{abstract}
Crank-Nicolson (C-N) Method combined with $ B $-spline basis set can be used in both time-dependent and time-independent calculations of the photoionization cross sections of atoms and molecules. For time-independent systems, it is found that C-N method in imaginary-time propagation (ITP) can converge directly to not only the ground state, but also excited and continuum states by controlling the time step size. The C-N method can also be directly applied in time-dependent calculations. Both time-dependent and time-independent calculations agree very well with previous results. This method can also be extended to two-electron systems directly. 
\end{abstract}
\maketitle

\section{Introduction}\label{I}
Photoionization of atoms and molecules is a fundamental physical process, which has been studied for many years.   
Accurate solution of continuum states is an essential task in the various quantum mechanical problems, such as the calculation of multiphoton ionization, above-threshold ionization (ATI), and electron scattering. However, the wavefunctions of continuum states oscillate to infinity. As a result, it is not as easy as the calculations of bound states. There are a number of methods to calculate the resonance states, such as complex scaling method, R-matrix method. However, they are not used to calculate continuum state with any positive energy. Although variational method \cite{Brosolo1,Brosolo2} and least-squares schemes \cite{Brosolo3} have been introduced in the calculation of continuum wavefunctions, efficient and accurate solutions are still necessary. In general, a variational approach can be used to calculate all the eigenstates of any quantum system using some appropriate basis set. After direct diagonalization of a Hamiltonian matrix for appropriate potentials, all the eigenvalues and wavefunctions are obtained. Negative eigenvalues correspond to bound states, while positive eigenvalues correspond to the continuum states. However, this method is not efficient for large scale matrix problems involving large basis sets for convergence and high accuracy. In any standard matrix diagonalization program, the demand on computer memory grows as $N^{2}$ and the CPU time as $N^{3}$ ($N$ is the size of the Hamiltonian matrix)\cite{Bian2}. When $N>10000$, the computation becomes very slow, even impractical in some cases. In addition, the obtained discrete continuum states may not be what we want because we don't know their boundary conditions. For certain chemical and physical problems, not all the eigenstates of a system are required. For example, to numerically solve time-dependent Schr\"{o}dinger equations (TDSE), one only needs the initial wavefunction, and then propagates it under a Hamiltonian including time-dependent interaction terms by using general split-operator methods (SOM)\cite{Herman,ADBSH}, Crank-Nicolson methods (C-N method)\cite{Crank,Sun}, or efficient large vector matrix operations\cite{Avila,Peng}. If the initial state in a simulation is the ground state, one does not need to diagonalize a basis set dependent Hamiltonian, but one propagates a TDSE for an arbitrary initial vector by an imaginary-time-propagation method (ITP method)\cite{ADB}. Complex time steps also allow to obtain higher order accuracy in SOM\cite{ADB}. In this paper, we show that ITP method based on C-N scheme can be generalized to converge to bound exited and unbound continuum states by changing time step. To our knowledge, this has not been reported. The ITP calculations and time-dependent calculations based on C-N method agree very well. 

The paper is organized as follows: The principle of ITP method for excited and continuum states is presented in Sec. \ref{II}. The application of the ITP method in photoionization of one-electron diatomic molecular ion H$_{2}^{+}$ and HeH$^{2+}$ is shown in Sec. \ref{III}. In Sec. \ref{IV}, we illustrate the application of the proposed ITP method in two-electron systems. The paper concludes in Sec. \ref{V}.

\section{Principle of ITP method}\label{II}
In this section, we briefly introduce next the principle of usual ITP method for ground state and our proposed ITP method based on C-N scheme for all the bound and continuum states.
\subsection{Usual ITP method for ground state}
 The principle of the ITP method can be found as follows. For a general TDSE (Atomic units, a.u., $e=\hbar=m_{e}=1$ are used unless otherwise specified),
\begin{equation}\label{E1}
  i\frac{\partial}{\partial t}|\Psi(\mathbf{r},t)\rangle=H|\Psi(\mathbf{r},t)\rangle,
\end{equation}
introducing the imaginary time $t\rightarrow -it$ in an ITP method, and if the Hamiltonian $H$ is time-independent, then the new time-dependent wavefunction after a time step $\Delta t$ can be written as:
\begin{equation}\label{E2}
 |\Psi(\mathbf{r},\Delta t)\rangle=\exp^{-H\Delta t}|\Psi(\mathbf{r},0)\rangle.
\end{equation}
Expanding an initial arbitrary wavefunction $|\Psi(\mathbf{r},0)\rangle $ as a linear combination of the time-independent eigenstates ${\varphi_{n}(\mathbf{r})}$ of $H$,
\begin{equation}\label{E3}
  |\Psi(\mathbf{r},0)\rangle =\sum_{n}a_{n}\varphi_{n}(\mathbf{r}),
\end{equation}
where $H\varphi_{n}(\mathbf{r})=E_{n}\varphi_{n}(\mathbf{r})$, then Eq.~(\ref{E2}) can be written as:
\begin{equation}\label{E4}
  |\Psi(\mathbf{r},\Delta t)\rangle=\sum_{n}\exp^{-E_{n}\Delta t}a_{n}\varphi_{n}(\mathbf{r}).
\end{equation}

\begin{figure}
\centering
\includegraphics [width=7cm,height=5.0cm, angle=0] {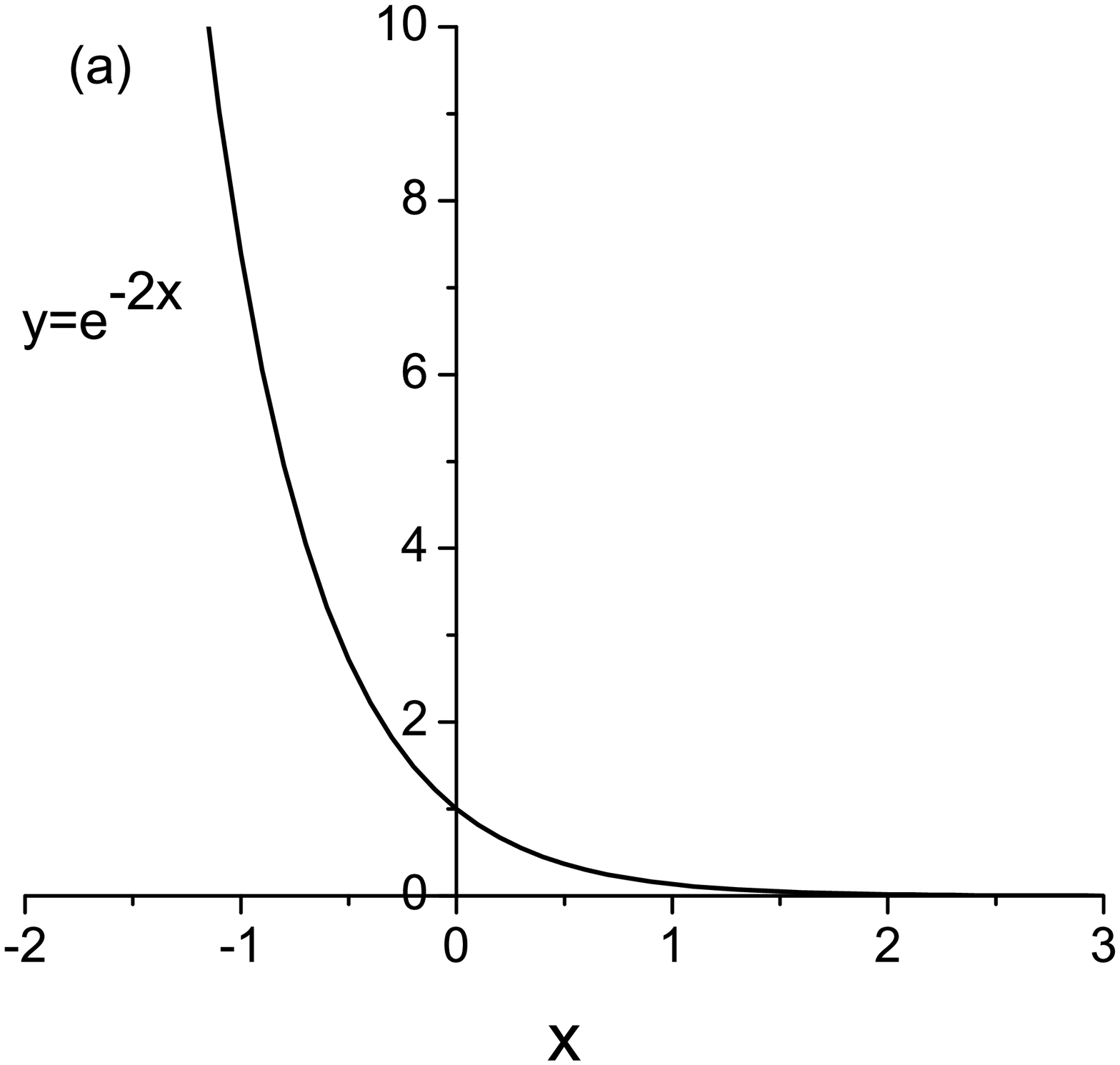}
\includegraphics [width=7cm,height=5.0cm, angle=0] {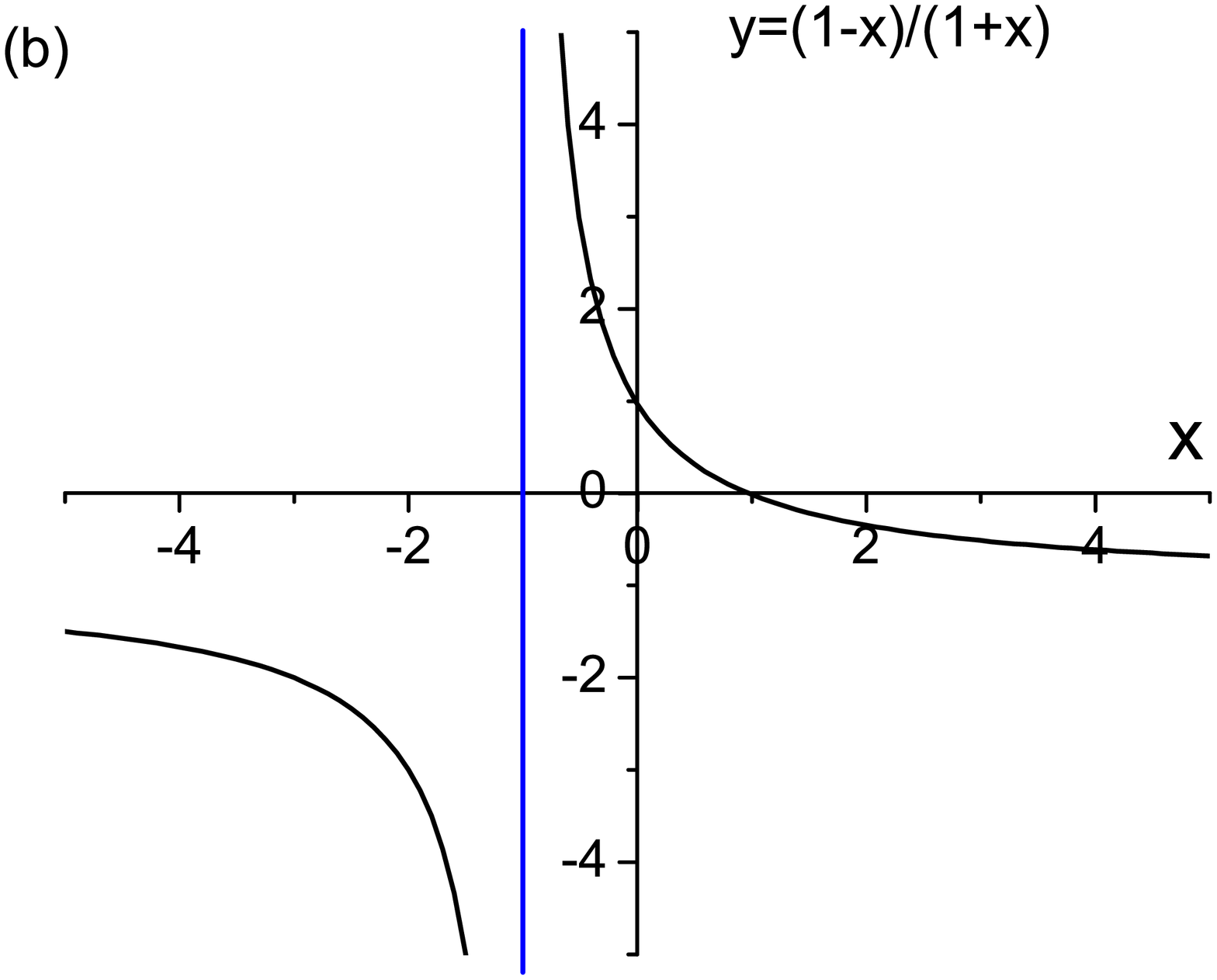}
     \caption{(Color online) Prefactors of Eq. (\ref{E4}) and Eq. (\ref{E6}) as a function of $x=E\Delta t/2$. }
 \label{Fig1}
\end{figure}

It is obvious that higher energy states will exponentially decay as a function of $E\Delta t$ (See Fig.~\ref{Fig1} (a)), at a rate which is faster than for the lower energy states. After sufficiently long time evolution, any arbitrary initial wavefunction will thus converge to the ground state. If we are interested in an excited state $\varphi_{n}(\mathbf{r})$, we have to at first obtain the converged ground state $\varphi_{1}(\mathbf{r})$ and other lower excited states $\varphi_{n^{\prime}}(\mathbf{r})$, with $n^{\prime}=2, 3, \cdots, n-1$, then project out all the lower energy states by $(1-\sum_{n^{\prime}=1}^{n-1}|\varphi_{n^{\prime}}(\mathbf{r})\rangle\langle\varphi_{n^{\prime}}(\mathbf{r})|)$ from the initial state before evolution\cite{Kosloff}. For higher excited states, this method becomes slowly convergent and difficult to apply. It is even not practical for continuum states due to the large degeneracy of those states.

\begin{table*}
  \centering
  \caption{Comparison of the eigenvalues of H $s$ states obtained by ITP method with time step $\Delta t$ and iterations $j$ and the analytic values $-1/2n^{2}$ (a.u.).}\label{T1}
\begin{ruledtabular}
\begin{tabular}{c c c c}
$j$& $E_{1}$ (a.u.) ($\Delta t=2$ a.u.)& $E_{2}$ (a.u.) ($\Delta t=10$ a.u.)& $E_{3}$ (a.u.) ($\Delta t=40$ a.u.)\\
\hline
1&  0.02074985099579411&0.01704393876353220&-0.03172899941642318\\
2& -0.004.535732379123197&-0.02707373759307351&-0.05502525262084053\\
3& -0.157925711875305&-0.102192312548041&-0.05553857185259704\\
4& -0.440683975870317&-0.122870665895126&-0.05555477488356365\\
5& -0.494240272539091&-0.124820611212683&-0.05555551614437183\\
6& -0.499235363892429&-0.124985275640302&-0.05555555352112400\\
7& -0.499870405769063&-0.124999310529914&-0.05555555544994874\\
8& -0.499975900374919&-0.125000203414494&-0.05555555555006548\\
9& -0.499995389821222&-0.125000111564971&-0.05555555555526998\\
10& -0.499999110747678&-0.125000039965192&-0.05555555555554072\\
11& -0.499999828045146&-0.125000012764312&-0.05555555555555482\\
12& -0.499999966722960&-0.125000003889258&-0.05555555555555550\\
13& -0.499999993558485&-0.125000001158381&-0.05555555555555556\\
14& -0.499999998752989&-0.125000000340924&-0.05555555555555559\\
15& -0.499999999758584&-0.125000000099687&-0.05555555555555550\\
16& -0.499999999953262&-0.125000000029042&-0.05555555555555559\\
17& -0.499999999990952&-0.125000000008444&-0.05555555555555552\\
18& -0.499999999998248&-0.125000000002452&-0.05555555555555557\\
19& -0.499999999999661&-0.125000000000712&-0.05555555555555549\\
20& -0.499999999999934&-0.125000000000207&-0.05555555555555556\\
21& -0.499999999999987&-0.125000000000060&-0.05555555555555559\\
22& -0.499999999999997&-0.125000000000017&-0.05555555555555559\\
23& -0.499999999999999&-0.125000000000005&-0.05555555555555555\\
24& -0.499999999999999&-0.125000000000002&-0.05555555555555554\\
25& -0.500000000000000&-0.125000000000000&-0.05555555555555555\\
$-1/2n^{2}$ &-0.5&-0.125&-0.0555555555555555(5)\\
\end{tabular}
\end{ruledtabular}
\end{table*}
\subsection{C-N ITP method for any bound state}

We show next that by using the C-N and ITP methods, we can simply restrict any initial arbitrary state to converge to desired excited states by changing the appropriate time step $\Delta t$ of the iteration. It can also be easily extended to calculate any continuum state, such as the final states in photoionization. The principle of C-N ITP method is similar to inverse iteration.

The C-N method is a unitary and unconditionally stable method\cite{Crank,Sun}. It is based on expressing the exponential operator $\exp^{-H\Delta t}$ to second-order accuracy in Eq. (\ref{E2}) as:
\begin{equation}\label{E5}
  \exp^{-H\Delta t}=\frac{1-H\Delta t/2}{1+H\Delta t/2}+O(\Delta t^{3}).
\end{equation}
Then Eq. (\ref{E4}) can be equivalently written for any arbitrary state,
\begin{equation}\label{E6}
    |\Psi(\mathbf{r},\Delta t)\rangle=\sum_{n}\frac{1-E_{n}\Delta t/2}{1+E_{n}\Delta t/2}a_{n}\varphi_{n}(\mathbf{r}).
\end{equation}
Setting $x=E_{n}\Delta t/2$, the factor in Eq. (\ref{E6}) is $\frac{1-x}{1+x}=1-\frac{2}{1+\frac{1}{x}}$, which does not evolve with exponential decay, but has a maximum around $x=-1$, as shown in Fig.~\ref{Fig1} (b). As a result, after long time evolution, only the eigenstate with $E_{n}\Delta t/2\approx-1$ remains dominant in the superposition of states in Eq. (\ref{E6}). However, in practice, since nondegenerate bound state energies are separated, $E_{n}\Delta t/2\approx-1$ is not a strict condition. We don't need to project out other lower bound states as described in the above subsection. As long as the absolute "action" distance $|E_{n}\Delta t/2+1|$ is smaller than that of other bound states, the state with energy $E_{n}$ will remain while other states decay faster. Choosing a time step $\Delta t$ which is positive and small enough, with $E_{1}\Delta t/2>-1$, or $\Delta t<2/|E_{1}|$, Eq. (\ref{E6}) will converge to the ground state with the lowest eigenvalue $E_{1}$. If one gradually next increases the time step $\Delta t$, when $E_{2}\Delta t/2+1<-1-E_{1}\Delta t/2$, then the ground state and other excited states excluding the first excited state will decay, thus converging to the first excited state. Similarly by further iteration, we can obtain other converged excited states directly only by gradually increasing the time step $\Delta t$, i.e., the process filters out all undesired states except one.

As a demonstration of the proposed method, we calculate the $s$ bound states of H. The wavefunction is expanded in radial $B$ splines\cite{Bachau,Bian1,Bian3} and angular spherical harmonics as:
\begin{equation}\label{E7}
   |\Psi(\mathbf{r},t)\rangle=\sum_{i=1}^{N}C_{i}(t)\frac{B_{i}^{k}(r)}{r}Y_{l}^{m}(\theta,\varphi),
\end{equation}
where $l=m=0$, $C_{i}(t)$ is the time-dependent coefficients of the corresponding $B$ splines, $k$ is the order of the $B$ splines.

The corresponding Hamiltonian is written as:
\begin{equation}\label{E8}
    H=-\frac{1}{2}\frac{d^{2}}{dr^{2}}+\frac{l(l+1)}{2r^{2}}-\frac{1}{r}.
\end{equation}

The time evolution at each step $\Delta t$ of the coefficients in Eq. (\ref{E7}) in the C-N method can be written as:
\begin{equation}\label{E9}
(\mathbf{S}+\mathbf{H}\Delta t/2)\mathbf{C}(\Delta t)=(\mathbf{S}-\mathbf{H}\Delta t/2)\mathbf{C}(0),
\end{equation}
where $\mathbf{S}$ and $\mathbf{H}$ are the overlap and Hamiltonian matrix\cite{Bian2,Bian3}, respectively, for Eq. (\ref{E8}) in the $B$-spline basis. Since $B$ splines are localized basis\cite{Bachau}, $\mathbf{S}$ and $\mathbf{H}$ are sparse band matrices, which greatly reduces the memory and computation time. In practical calculations, the right side of Eq. (\ref{E9}) is a matrix vector multiplication. Then $\mathbf{C}(\Delta t)$ can be obtained by solving sparse linear equations.

In our calculation, the radial space dimension is truncated at $r_{max}=100$ a.u.; the wavefunction is expanded by $N=$200 $B$ splines with order $k=7$. The initial coefficients are set as $C_{i}(0)=1/\sqrt{N}$, for $i=1, 2, \ldots, N$.
As illustrated in Table~\ref{T1}, with a time step $\Delta t=2$ a.u., which satisfies $\Delta t<2/|E_{1}|$, convergence to the ground state is obtained readily with energy $E_{1}=-0.5$ a.u. with machine accuracy after 25 iterations. When we increase the time step to $\Delta t=10$ a.u., which satisfies the condition $E_{2}\Delta t/2+1<-1-E_{1}\Delta t/2$ as discussed above, the first excited state $n=2$ is obtained. Similarly, other higher excited states can be easily obtained directly by controlling the time step $\Delta t$. This method is not restricted to previous known systems, but it is a general method as shown next.

\subsection{C-N ITP method for continuum states}

\begin{figure}
\centering
\includegraphics [width=9cm,height=7.0cm, angle=0] {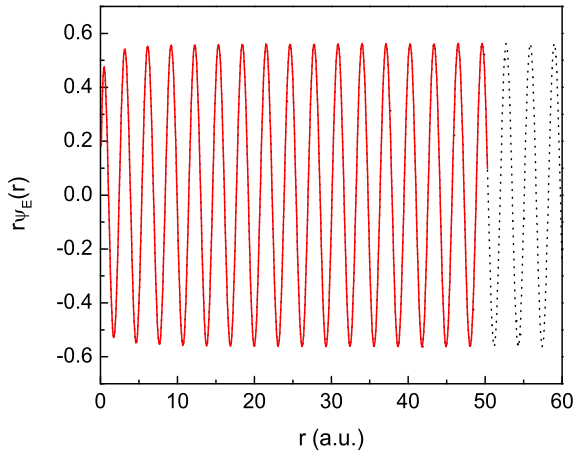}

     \caption{(Color online) Comparison of Hydrogen continuum radial state $r\Psi_{E}(r)$ obtained by ITP method (with iterations $j=18$) (solid line) with the analytic Coulomb wave function (dotted line) with $l=0$, $E=2$ a.u. Both are normalized on the energy scale.}
 \label{Fig2}
\end{figure}

For field-free continuum states with positive energy $E_{c}$, except for the direct diagonalization\cite{Bachau}, there is a least-squares method\cite{Fischer,Madsen} to obtain $E_{c}$ and $\Psi_{c}$. This is based on the idea of minimizing
the residual vector $(H-E)\Psi$. We show next that we can directly extend the ITP method to obtain any continuum state. As shown above, the ITP method converges an arbitrary vector to the bound eigenstate $\varphi_{n}$ at the action value $E_{n}\Delta t/2\approx-1$, or energy $E_{n}\approx-2/\Delta t$. If we want the continuum states with $E_{c}>0$, we only have to change the sign of the time step $\Delta t\rightarrow -\Delta t$ in the above restriction; or equivalently, we transform the TDSE in Eq.(\ref{E1}) by a new ITP method with $t\rightarrow it$. Then the arbitrary initial state converges to continuum state $\Psi_{c}$ with positive energy $E_{c}\approx 2/\Delta t$. For bound states, we don't know the energies and wavefunctions, but the boundary condition $\varphi_{n}(r_{max})=0$ is known. However, for continuum state, its energy is known, we don't know the boundary condition, the above process will not exactly converge to $E_{c}$. Since the wavefunctions of continuum states oscillate to infinity, there must be a series of points $r_{c}$ with $\Psi_{c}(r_{c})=0$ \cite{Mercouris,Mercouris2}. In practice, we can safely set the boundary condition $\Psi_{c}(r_{max})=0$ in our calculations. We smoothly change the size of the box $r_{max}$ in the ITP method until the expected eigenvalue $\langle\Psi|H|\Psi\rangle$/$\langle\Psi|\Psi\rangle$ matches the desired eigenvalue $E_{c}$ with required accuracy\cite{Bachau}, then we can say $r_{max}=r_{c}$. To calculate the continuum state with energy $E_{c}=2$ a.u., we propagate an arbitrary vector with a time step $\Delta t=1$ a.u.  The obtained wavefunction after $j=$18 iterations is illustrated in Fig.~\ref{Fig2}, which is normalized on the energy scale. We also plot the exact regular Coulomb wave function as a comparison in Fig.~\ref{Fig2}. The illustrated result clearly shows that this new ITP method for obtaining continuum states is very accurate.

\section{Applications of ITP method in photonionization of one-electron diatomic molecules}\label{III}
Molecular systems are more complex than atoms. There are no simple exact analytic solutions even for the simplest one-electron molecular ion H$_{2}^{+}$\cite{Bates,mad}.  Although a Green's function in spheroidal coordinates can be obtained by the quantum defect method\cite{Davy,Kawai}, accurate numerical solutions are very useful, especially for the TDSE\cite{Madsen}. In this section we will show that our ITP method can be directly extended to diatomic molecular systems.

We use prolate spheroidal coordinates ($\xi$,$\eta$,$\phi$) to represent the electron. $\phi$ is the azimuthal angle, the internuclear distance is $R$. If $r_{1}$ and $r_{2}$ is the distance of the electron from the two nuclei, we have
\begin{equation}\label{E10}
\xi=(r_{1}+r_{2})/R,\ \eta=(r_{1}-r_{2})/R,
\end{equation}
with $\xi\in[1,\infty]$, $\eta\in[-1,1]$.
The field-free Hamiltonian in the fixed-nuclei approximation is\cite{Bian1}:
\begin{eqnarray}\label{E11}
  H_{0}&=&-\frac{1}{2}\nabla^{2}+V \nonumber\\
   &=&  -\frac{2}{R^{2}(\xi^{2}-\eta^{2})}\Bigg[\frac{\partial}{\partial\xi}\left((\xi^{2}-1)\frac{\partial}{\partial\xi}\right)\nonumber\\
    &&+\frac{\partial}{\partial\eta}\left((1-\eta^{2})\frac{\partial}{\partial\eta}\right)
   +\left(\frac{1}{\xi^{2}-1}+\frac{1}{1-\eta^{2}}\right)\frac{\partial^{2}}{\partial\phi^{2}}\Bigg]\nonumber\\
   &&-2\frac{(Z_{1}+Z_{2})\xi-(Z_{1}-Z_{2})\eta}{R(\xi^{2}-\eta^{2})},
\end{eqnarray}
where $Z_{1}$ and $Z_{2}$ are the nuclear charges. We expand the electronic wave function in a $B$-spline basis\cite{Bian1} as follow:
\begin{eqnarray}\label{E12}
  &&\psi(\xi,\eta,\varphi)  \nonumber \\
   && =\sum_{i,j,m}C_{i,j}^{m}(\xi^{2}-1)^{\frac{|m|}{2}}B_{i}(\xi)(1-\eta^{2})^{\frac{|m|}{2}}B_{j}(\eta)\frac{e^{im\phi}}{\sqrt{2\pi}}.\nonumber
\end{eqnarray}

The volume element is $d\tau=(R/2)^{3}(\xi^{2}-\eta^{2})d\xi d\eta d\phi$. For H$_{2}^{+}$, $Z_{1}=Z_{2}=1$. We have calculated the bound state energy with $m=0$, $R=2$ a.u. The $\eta$ dimension is expanded into 20 B splines with order 7. The $\xi$ direction is truncated at $\xi_{max}=80$ a.u., and is expanded by 80 B splines with order 7. The results with iterations $j=8$ are presented in Table~\ref{T2} as a comparison with Ref.\cite{mad}. Our present method with a small number of iterations $j$ or total time $t=j\Delta t$ is very accurate with a small basis and faster to converge to the desired states than usual methods.
\begin{table}
  \centering
  \caption{Comparison of the $m=0$ eigenvalues of H$_{2}^{+}$ obtained by ITP method and the values in Ref.\cite{mad}. The internuclear distance is $R=2$ a.u., the iterations $j=8$ in ITP method.}\label{T2}
\begin{ruledtabular}
\begin{tabular}{c c c c}
state & $\Delta t$ (a.u.) & $E$ (a.u.) (ITP) & $E$ (a.u.) in Ref.\cite{mad} (a.u.)\\
\hline
1$\sigma_{g}$&2&-1.1026342144949&-1.1026342144949\\
1$\sigma_{u}$&3&-0.6675343922023&-0.6675343922024\\
2$\sigma_{g}$&6&-0.3608648753394&-0.3608648753383\\
2$\sigma_{u}$&8&-0.2554131650864&-0.2554131650857\\
\end{tabular}
\end{ruledtabular}
\end{table}

The two-center wavefunction is also separable as:
\begin{equation}\label{E13}
   \psi(\xi,\eta,\phi)=U(\xi)V(\eta)\Phi(\phi),
\end{equation}
where $\Phi(\phi)=\frac{e^{im\phi}}{\sqrt{2\pi}}$, and $U(\xi)$ and $V(\eta)$ satisfy separate differential equations,
\begin{equation}\label{E14}
   \frac{d}{d\xi}[(\xi^{2}-1)\frac{dU}{d\xi}]+[-A_{mq}+c^{2}(\xi^{2}-1)+a\xi-\frac{m^{2}}{\xi^{2}-1}]U=0,
\end{equation}
\begin{equation}\label{E15}
   \frac{d}{d\eta}[(1-\eta^{2})\frac{dV}{d\eta}]+[A_{mq}+c^{2}(1-\eta^{2})+b\eta-\frac{m^{2}}{1-\eta^{2}}]V=0,
\end{equation}
where $A_{mq}$ is the separation constant, $a=R(Z_{1}+Z_{2})$, $b=R(Z_{2}-Z_{1})$, $c^{2}=R^{2}E/2$.

\begin{table}
  \centering
  \caption{One photon partial absorbtion cross sections (For comparison, Eq.~(\ref{E17}) is divided by 3.) from the H$_{2}^{+}$ ground state 1$\sigma_{g}$ ($q=0$). The internuclear distance is $R=2$ a.u., 1Mb=10$^{-18}$ cm$^{2}$.}\label{T3}
\begin{ruledtabular}
\begin{tabular}{c c c }
Photoelectron energy & $p\sigma_{u}$ ($q$=1)  &$f\sigma_{u}$ ($q$=3) \\
$E_{c}$ (a.u.)&($\times10^{-2}$ Mb)&($\times10^{-2}$ Mb)\\
\hline

1&0.688$^{a}$&0.907$^{a}$\\
&0.694$^{b}$& 0.904$^{b}$\\
&0.54$^{c}$&0.94$^{c}$\\
2&0.517$^{a}$&0.618$^{a}$\\
&0.516$^{b}$&0.618$^{b}$\\
4&0.202$^{a}$&0.220$^{a}$\\
&0.200$^{b}$&0.222$^{b}$\\
10&0.0140$^{a}$&0.00192$^{a}$\\
&0.0138$^{b}$&0.00190$^{b}$\\
\end{tabular}
\end{ruledtabular}
$^{a}$Present work, $^{b}$Ref.\cite{Richards}, $^{c}$ Ref.\cite{Bates}
\end{table}

\begin{figure}
\centering
\includegraphics [width=9cm,height=7.0cm, angle=0] {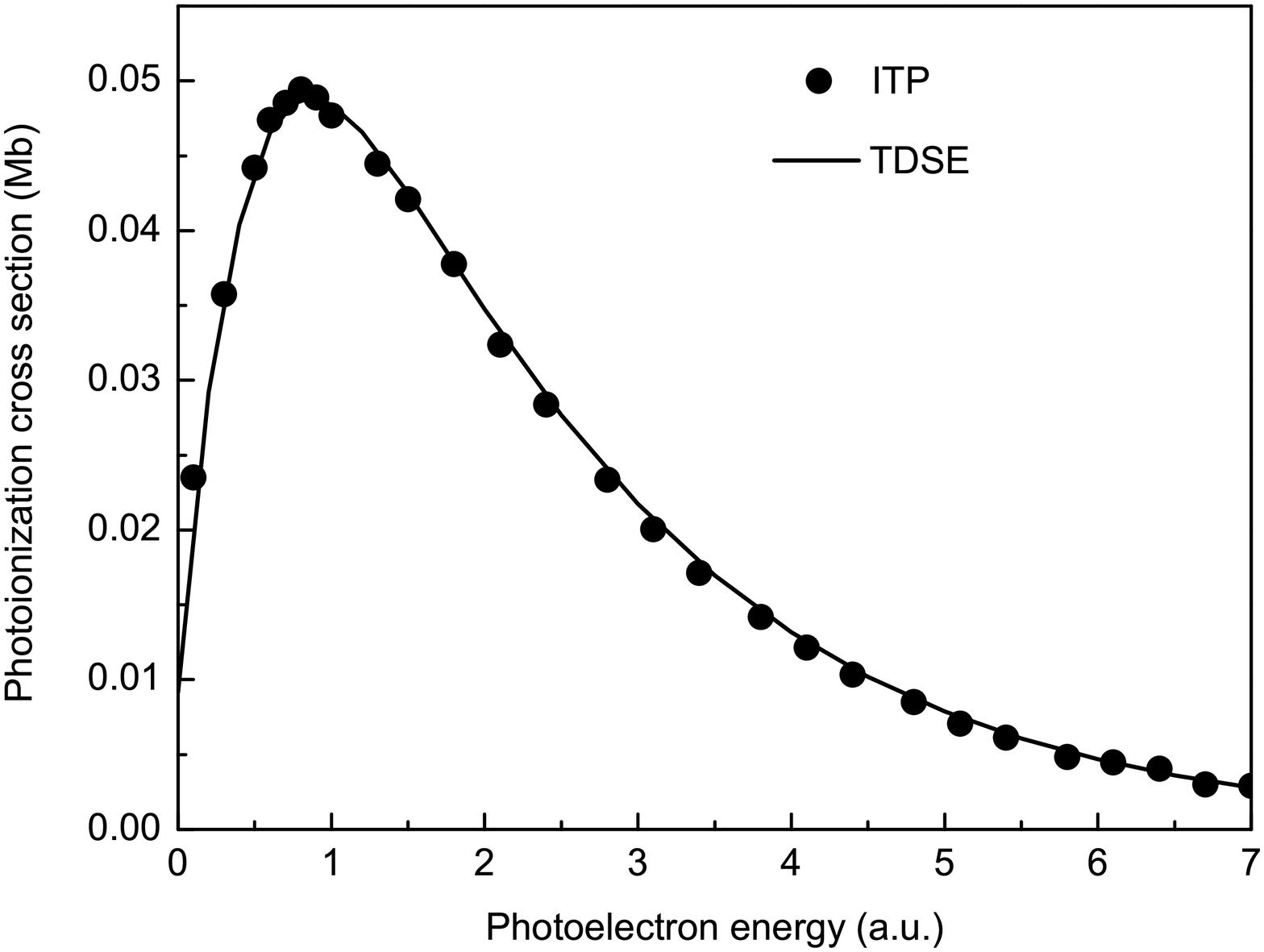}
\caption{One-photon ionization cross sections for H$_{2}^{+}$ as a function of photoelectron energy. The initial state is the ground 1$\sigma_{g}$ state. The laser polarization is parallel to the molecular axis. The internuclear distance $R=2$ a.u. The time-dependent results (solid line) are compared with those by ITP method (circles).}
 \label{Fig3}
\end{figure}

\begin{figure}
\centering
\includegraphics [width=9cm,height=7.0cm, angle=0] {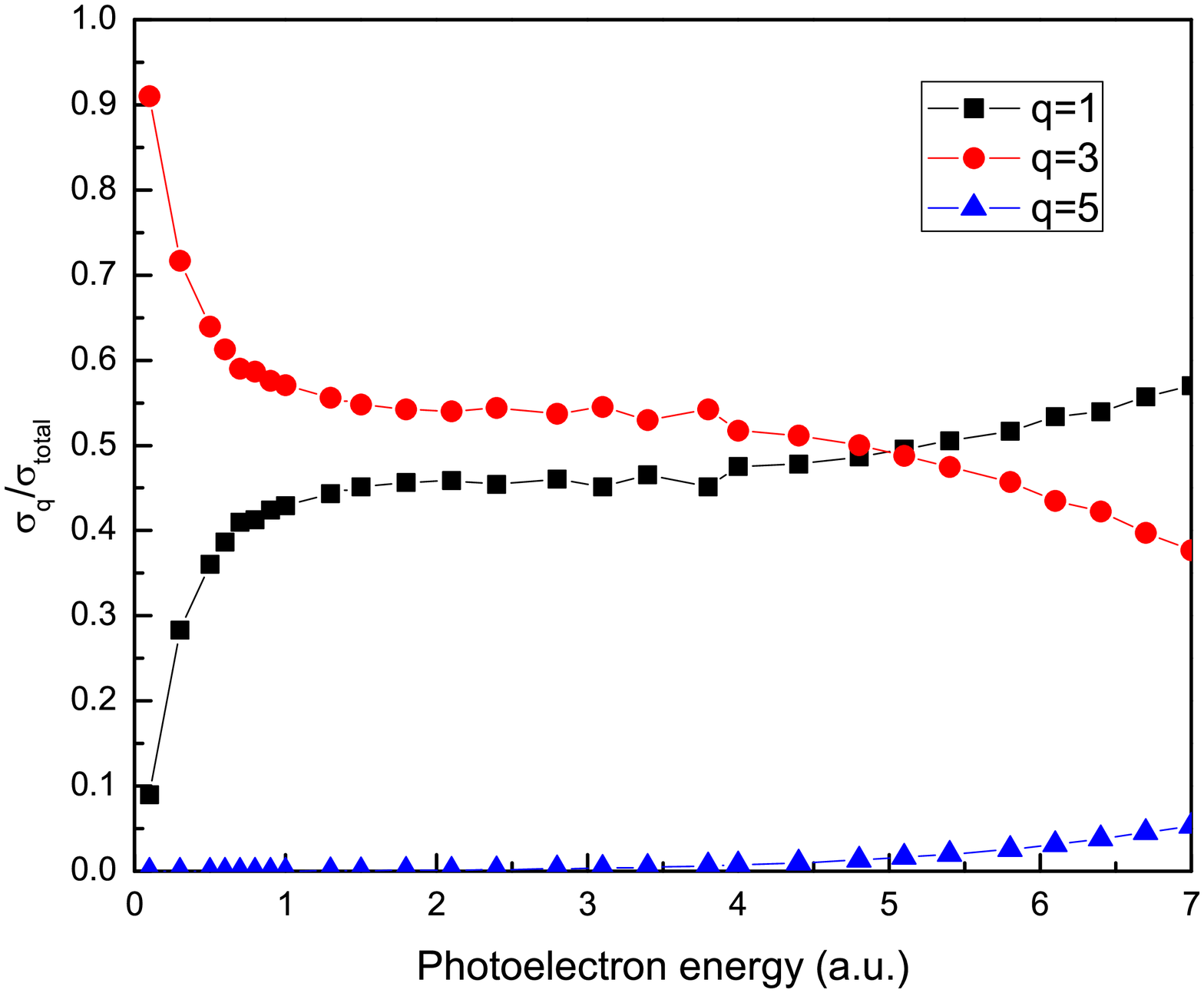}
\caption{(Color online) Relative contribution of each partial cross section $\sigma_{q}$ to the total one-photoionization cross section $\sigma_{total}$. The initial state is the ground 1$\sigma_{g}$ state.}
 \label{Fig4}
\end{figure}

\begin{figure}
\centering
\includegraphics [width=9cm,height=7.0cm, angle=0] {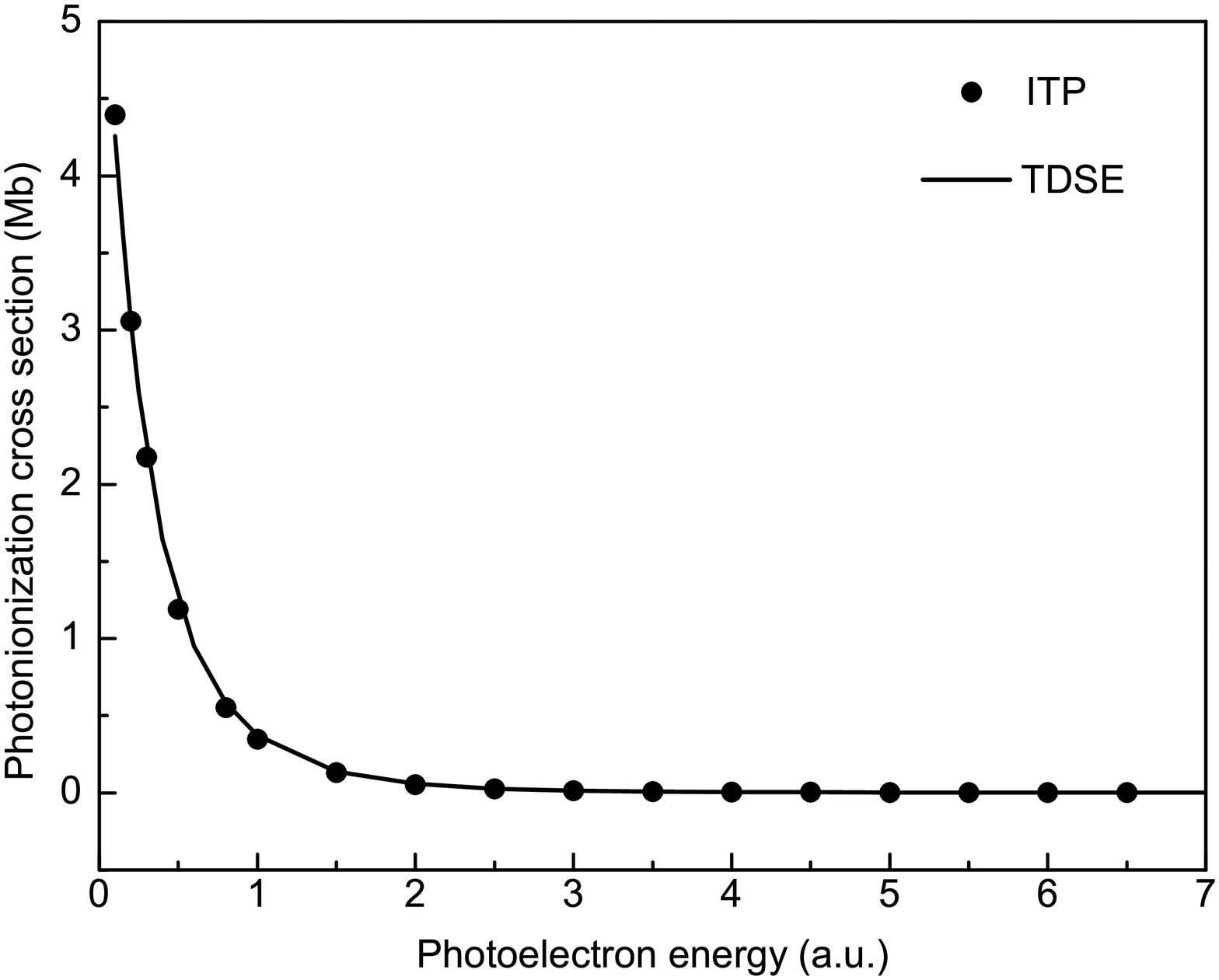}
     \caption{Same as Fig. \ref{Fig3}}, but the initial state is the first excited state 1$\sigma_{u}$.
 \label{Fig5}
\end{figure}

\begin{figure}
\centering
\includegraphics [width=9cm,height=7.0cm, angle=0] {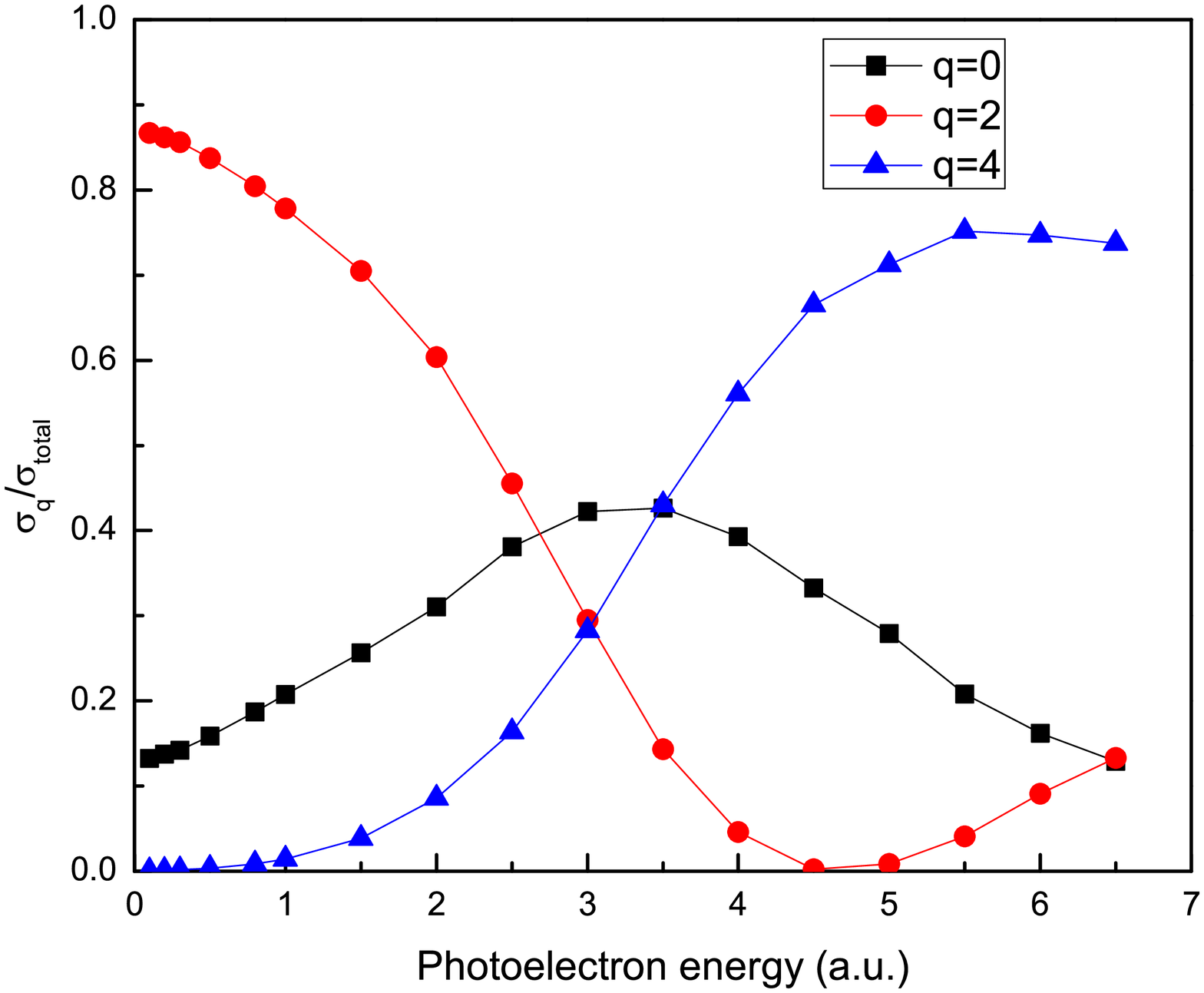}
     \caption{(Color online) Same as Fig. \ref{Fig4}}, but the initial state is the first excited state 1$\sigma_{u}$.
 \label{Fig6}
\end{figure}

For a continuum state with energy $E_{c}$, $A_{mq}$ (where $q$ is equal to the number of zeros of the function $V(\eta)$) can be calculated by expanding $V(\eta)$ in a series of associated Legendre polynomials\cite{Ponomarev}. It can also be obtained by a diagonalization of Eq. (\ref{E15}) \cite{HBachau} or using the ITP method introduced above. In the present work, we expand $V(\eta)$ into 20 $B$ splines with order 7, and use the diagonalization method to have the separation constant $A_{mq}$ and the corresponding $V(\eta)$. To calculate the radial wavefunction $U(\xi)$ in Eq. (\ref{E14}), we use the ITP method introduced above. Instead of fitting $E_{c}$, we smoothly vary the box $\xi_{max}$ to fit the separation constant $A_{mq}$ obtained from Eq. (\ref{E15}). The normalization of the radial function is realized by fitting the asymptotic behavior of $U(\xi)$ at $\xi\gg1$\cite{Guan,Guan2,Miyagi}:
\begin{equation}\label{E16}
   U(\xi)\rightarrow\frac{1}{R\xi}\sqrt{\frac{8}{k \pi}}\sin(\frac{kR\xi}{2}+\frac{2}{k}\ln(kR\xi)-\frac{l\pi}{2}+\Delta),
\end{equation}
where $\Delta$ is the phase of the radial function.

Once we have the continuum state, in examples discussed in the introduction, one can calculate general one-photon absorbtion partial cross sections given by\cite{HBachau}:
\begin{equation}\label{E17}
  \sigma_{q}=4\pi^{2}\alpha\omega|\langle i|D|f \rangle|^{2},
\end{equation}
where $|i\rangle$ and $|f\rangle$ are the initial and final state, respectively. $D$ is the transition operator $z$, which is $R\xi\eta/2$ in the length gauge. $\alpha$ is the fine-structure constant, and $\omega$ is the photon energy.

Table~\ref{T3} shows the one-photon partial absorbtion cross sections from the H$_{2}^{+}$ ground state at equilibrium internuclear distance $R=2$~a.u. to different continuum states $|f\rangle=|E_{c}\rangle$, where $E_{c}=h\nu-I_{p}$.  Again, using the new ITP method, our results agree well with Ref.\cite{Richards} from low to very high energy. The relative difference is less than 1\%. For the laser field polarization parallel to the molecular axis, we present the total one-photon ionization cross section $\sigma_{total}$ as a function of photoelectron energy $E_{c}$ in Fig.~\ref{Fig3}. It agrees well with the results by complex scaling method\cite{Tao}. The total cross sections have a maximum around $E_{c}=0.8$ a.u. We also present the relative contribution of each partial wave to the total cross section in Fig.~\ref{Fig4}. Because of the symmetry, only the transitions to states with different parities are allowed. For lower photoelectron energy, the transition to $q=3$ states is dominant. The contributions from $q=1$ and $q=3$ states are comparable in high energy region. The contribution of $q=5$ states is almost negligible.

C-N method can also be used in real-time TDSE. For very short time step $\Delta t$, Eq. (\ref{E9}) can be written as:
\begin{equation}\label{E18}
\mathbf{C}(t+\Delta t)=\frac{\mathbf{S}-i\mathbf{H}(t+\Delta t/2)\Delta t/2}{\mathbf{S}+i\mathbf{H}(t+\Delta t/2)\Delta t/2}\mathbf{C}(t).
\end{equation}

A sine-squared laser pulse with duration $\tau$ ten optical cycles with different photon frequencies are used in the calculation of TDSE. The intensity is $I=1\times10^{13}$ W/cm$^{2}$. At the end of the pulse $t_{f}$, the one-photon ionization cross section is given by\cite{Colgan}:
\begin{equation}\label{E19}
\sigma_{total}=\left(\frac{\omega}{I}\right)\frac{P}{T_{eff}}.
\end{equation}
where $T_{eff}=\frac{3}{8}\tau$, and $P=1-\sum_{n}|\langle\varphi_{n}|\Psi(t_{f})\rangle|^{2}$. The TDSE results are also presented in Fig.~\ref{Fig3}. One can see that the total one-photon ionization cross sections by ITP method agree well those by TDSE.

We also calculated the one-photon ionization cross sections from the first excited state 1$\sigma_{u}$ ($q$=1) of H$_{2}^{+}$. The results by ITP and TDSE methods are shown in Fig.~\ref{Fig5}. They also agree well with each other, showing the accuracy of our numerical methods. The total cross sections are quite larger than those from the ground state due to its lower ionization potential, but they have a fast decay as the increase of the photoelectron energy. The contributions of each partial wave are illustrated in Fig.~\ref{Fig6}. As the increase of photoelectron energy, the contribution from $q=0$ states has a maximum around $E_{c}=3$ a.u.; the contribution from $q=2$ states has a fast decay with $E_{c}<4.5$ a.u., then increases; while the contribution from $q=4$ states gradually increases.

For the simplest asymmetric one-electron molecular ion HeH$^{2+}$, it has demonstrated a number of new features different from H$_{2}^{+}$ in high-order harmonic generation (HHG) \cite{Bian1,Bian4} and enhanced ionization (EI)\cite{Kamta,Kamta1}. We show next the application of ITP method in HeH$^{2+}$ nonsymmetric system.

\begin{figure}
\centering
\includegraphics [width=9cm,height=7.0cm, angle=0] {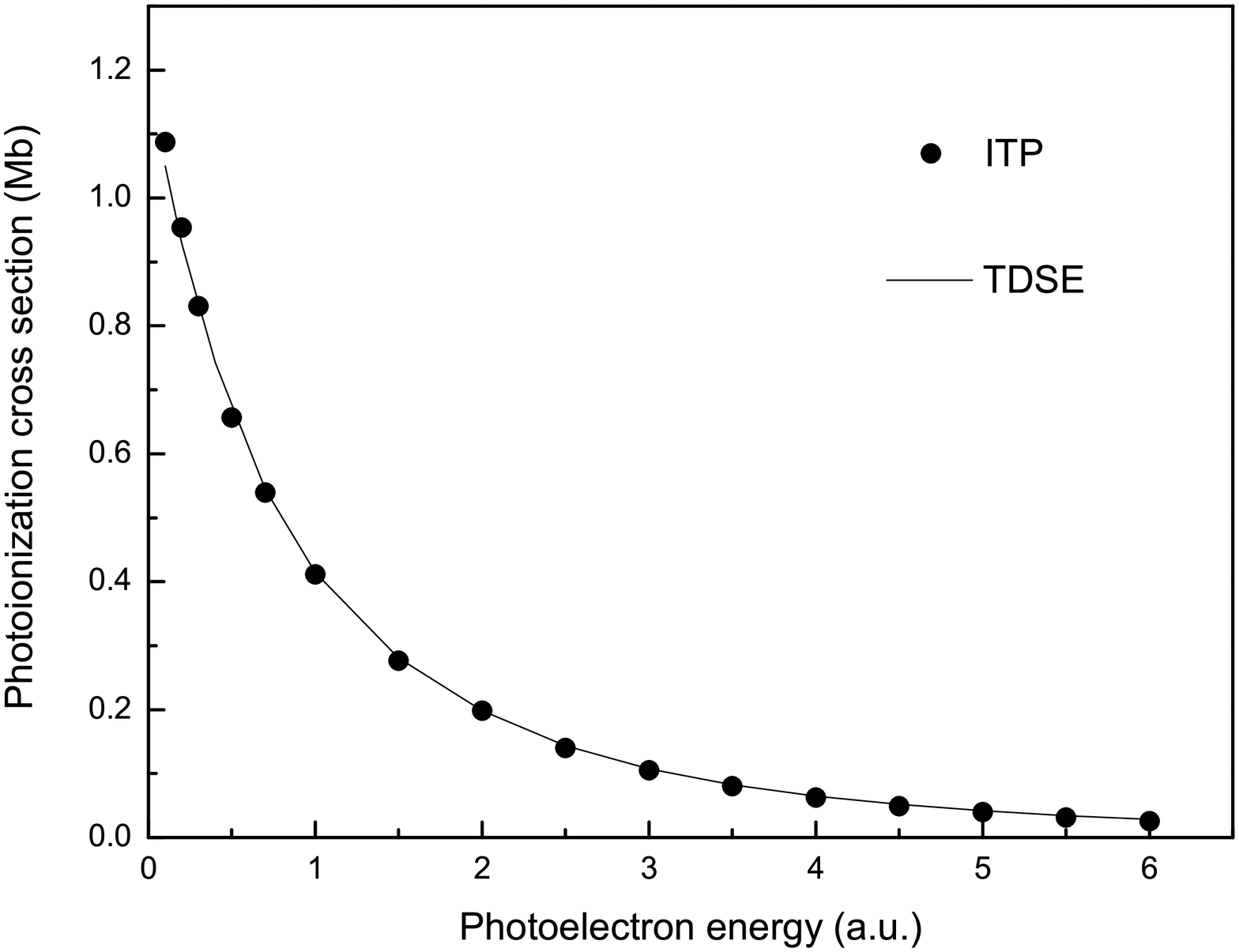}
\caption{One-photon ionization cross sections for HeH$^{2+}$ as a function of photoelectron energy. The initial state is the ground 1$s\sigma$ state. The laser polarization is parallel to the molecular axis. The internuclear distance $R=4$ a.u. The time-dependent results (solid line) are compared with those by ITP method (circles).}
 \label{Fig7}
\end{figure}

\begin{figure}
\centering
\includegraphics [width=9cm,height=7.0cm, angle=0] {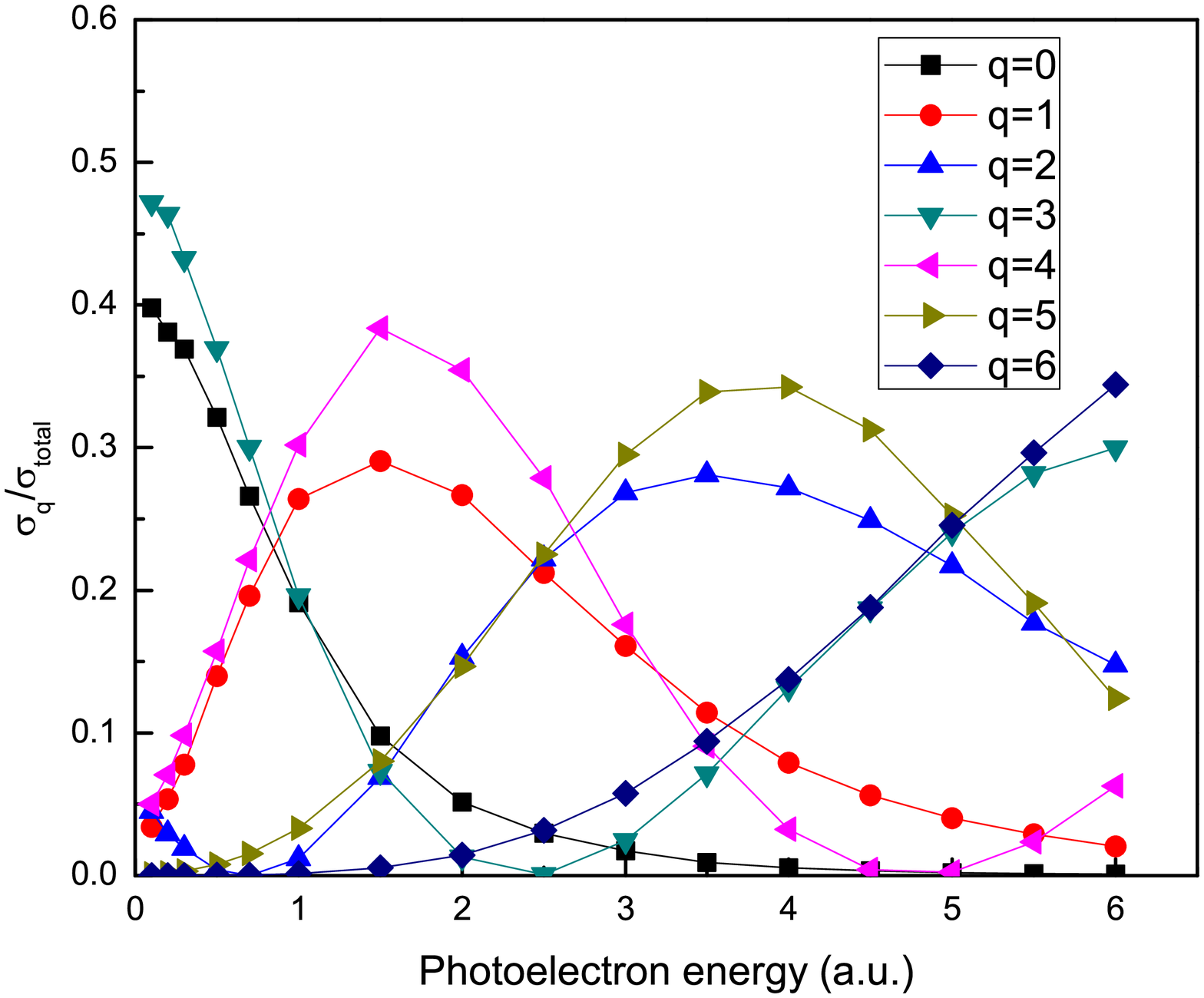}
\caption{(Color online) Relative contribution of each partial cross section $\sigma_{q}$ to the total one-photoionization cross section $\sigma_{total}$. The initial state is the ground 1$s\sigma$ state.}
 \label{Fig8}
\end{figure}

We fix the internuclear distance $R=4$ a.u. (the equilibrium internuclear distance is around $R=3.89$ a.u.). The eigenvalues of lower bound states obtained by ITP method are presented in Table~\ref{T4}. They agree well with recent work.\cite{Campos} The one-photon ionization cross sections from the ground 1$s\sigma$ state of HeH$^{2+}$ as a function of photoelectron energy by ITP and TDSE methods are illustrated in Fig. \ref{Fig7}. They agree very well and demonstrate a fast decay as a function of photoelectron energy. Due to the loss of symmetry, the final states may contain any partial wave. We expand the final wave with $q_{max}=6$ in the ITP method. The contribution of each partial wave is presented in Fig. \ref{Fig8}. For $q=0$, its contribution decays fast as the increase of photoelectron energy. The contributions of $q=1$ and $q=4$ states have a maximum around $E_{c}=1.5$ a.u., while the contributions of $q=2$ and $q=5$ states have a maximum around $E_{c}=4$ a.u. The contribution of $q=3$ state has a minimum around $E_{c}=2.5$ a.u. The contribution of $q=6$ state gradually increase as a function of photoelectron energy. Although the relative contributions of each partial wave is complex, the ITP method allows us extracting the information of each partial wave directly, which is important to calculate the angular distribution of photoelectrons\cite{Guan,Madsen}.

\begin{figure}
\centering
\includegraphics [width=9cm,height=7.0cm, angle=0] {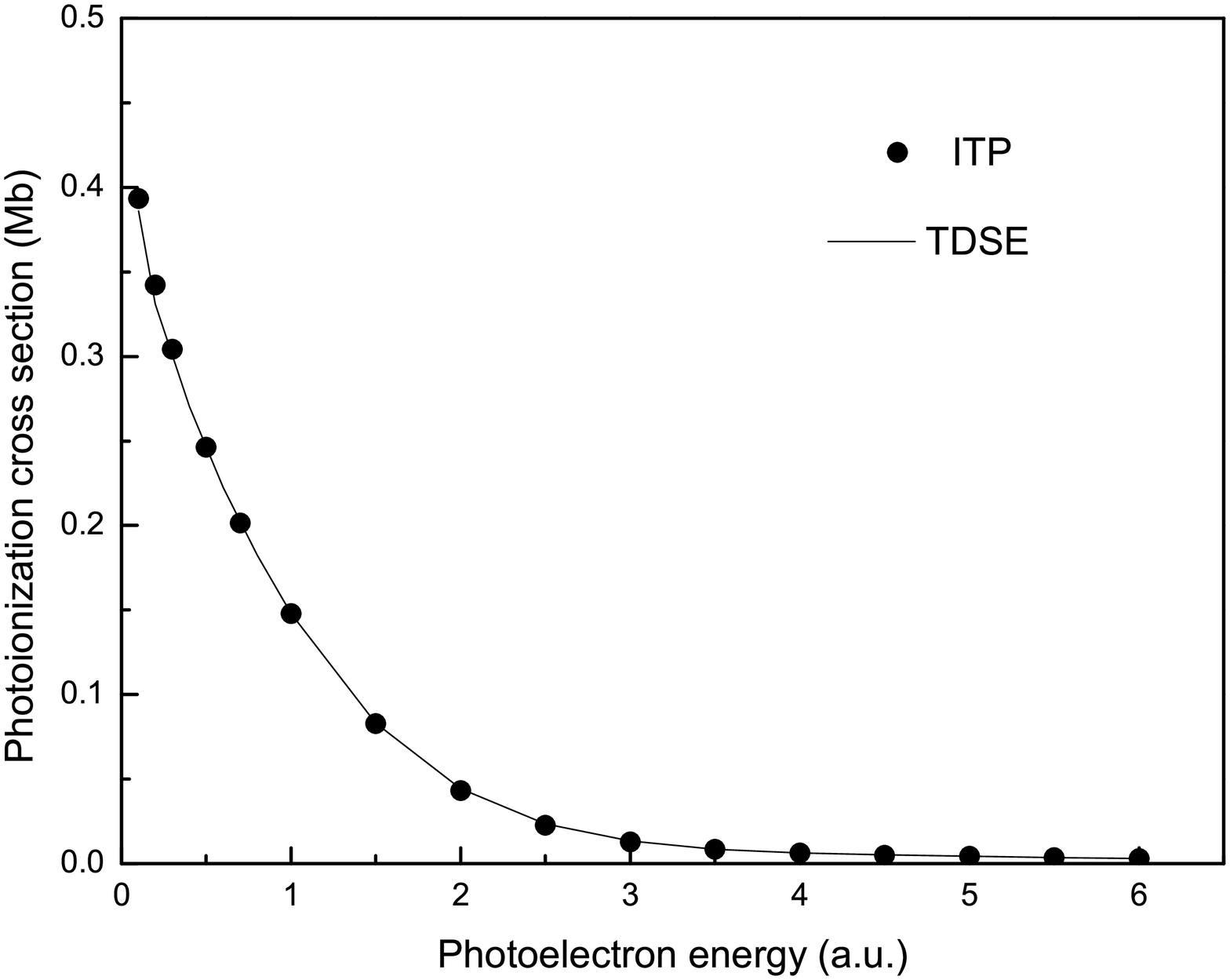}
     \caption{Same as Fig. \ref{Fig7}}, but the initial state is the first excited state 2$p\sigma$.
 \label{Fig9}
\end{figure}

\begin{figure}
\centering
\includegraphics [width=9cm,height=7.0cm, angle=0] {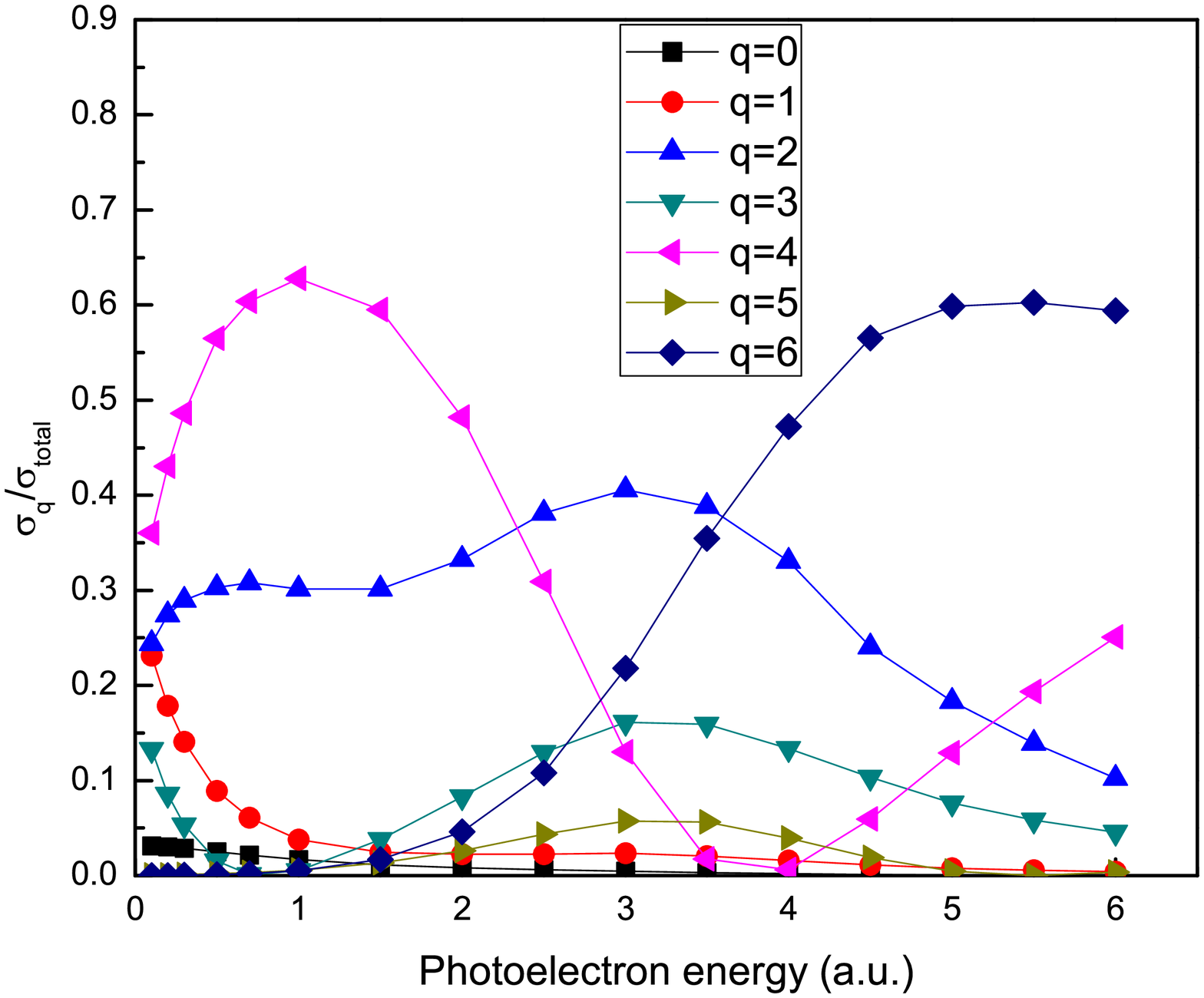}
     \caption{(Color online) Same as Fig. \ref{Fig8}}, but the initial state is the first excited state 2$p\sigma$.
 \label{Fig10}
\end{figure}

The one-photon ionization cross sections from the first excited state 2$p\sigma$ of HeH$^{2+}$ as a function of photoelectron energy by ITP and TDSE methods are illustrated in Fig. \ref{Fig9}. It is similar to that of 1$s\sigma$ state, presenting a fast decay as the increase of photoelectron energy. The relative contribution of each partial wave shown in Fig.~\ref{Fig10} is more complex than that in Fig.~\ref{Fig8}. The main contributions are from $q=2,3,4,6$ states.

\begin{table}
  \centering
  \caption{Comparison of the $m=0$ eigenvalues of HeH$^{2+}$ obtained by ITP method and the values in Ref.\cite{Campos}. The internuclear distance is $R=4$ a.u., the iterations $j=8$ in ITP method.}\label{T4}
\begin{ruledtabular}
\begin{tabular}{c c c c}
state & $\Delta t$ (a.u.) & $E$ (a.u.) (ITP) & $E$ (a.u.) in Ref.\cite{Campos} (a.u.)\\
\hline
1$s\sigma$&1&-2.250605387820&-2.250605387827\\
2$p\sigma$&2&-1.031081311774&-1.031081311774\\
2$s\sigma$&3&-0.6809853203162&-0.680985320316\\
3$p\sigma$&4&-0.4493213894454&-0.449321389445\\
\end{tabular}
\end{ruledtabular}
\end{table}

\section{Applications of ITP method in two-electron systerms}\label{IV}
It is very important to accurately solve the many-electron systems to study the electronic correlation effects. In this section, we extend our method to be used in two-electron atomic systems\cite{ZhaL}. The non-relative Hamiltonian in Eq.(\ref{E1}) can be written as:
\begin{equation}\label{E20}
H=\sum_{i=1}^{2}\left(\frac{\hat{\mathbf{p}}^{2}_{i}}{2}-\frac{Z}{r_{i}}\right)+\frac{1}{|\mathbf{r}_{1}-\mathbf{r}_{2}|},
\end{equation}
where $Z$ is the nuclear charge. The total electronic electronic wavefunction is expanded in coupled spherical harmonics:
\begin{eqnarray}\label{E21}
   &&|\Psi(\mathbf{r}_{1},\mathbf{r}_{2},t)\rangle \nonumber\\
   && =\sum_{L,M}\sum_{l_{1},l_{2}}\frac{\psi_{l_{1}l_{2}}^{LM}(r_{1},r_{2},t)}{r_{1} r_{2}}\sum_{m_{1},m_{2}}C_{m_{1}m_{2}{M}}^{l_{1}l_{2}L}Y_{m_{1}}^{l_{1}}(\Omega_{1})Y_{m_{2}}^{l_{2}}(\Omega_{2}), \nonumber\\
\end{eqnarray}
where $C_{m_{1}m_{2}{M}}^{l_{1}l_{2}L}$ is a Clebsch-Gordan coefficient, and $\psi_{l_{1}l_{2}}^{LM}(r_{1},r_{2},t)$ is expanded by $B$ splines as:
\begin{equation}\label{E22}
 \psi_{l_{1}l_{2}}^{LM}(r_{1},r_{2},t)=\sum_{i,j} C^{l_{1},l_{2},LM}_{i,j}(t)B_{i}(r_{1})B_{j}(r_{2}).
\end{equation}

The Hamiltonian matrix can be obtained easily except the two-electron integrals. We expand the interelectron repulsion term by a multipole expansion:
\begin{equation}\label{E23}
   \frac{1}{|\mathbf{r}_{1}-\mathbf{r}_{2}|}=\sum_{l}\frac{(r_{1},r_{2})_{<}^{l}}{(r_{1},r_{2})_{>}^{l+1}}P_{l}(\cos\theta_{12}),
\end{equation}
where $P_{l}(\cos\theta_{12})$ is a Legendre function. The angular integral can be done analytically, which can be found in Ref.\cite{Pindzola}. The radial integral is based on Poisson's equation, which is introduced in Ref.\cite{McCurdy}. Using the ITP method presented above, we calculated the S states of He. The radial space is truncated at $r_{max}=40$ a.u., and the radial wavefunction in Eq.(\ref{E22}) is expanded by 60 $B$ splines with order 8. The angular part in Eq.(\ref{E21}) is expanded by $l_{1,max}=l_{2,max}=10$. The two-electron integral in in Eq.(\ref{E23}) is truncated with $l_{max}=35$. The obtained singlet and triplet eigenenergies are presented in Table~\ref{T5} with a comparison of the values in Ref.\cite{Drake}. In spite of small size of basis, the accuracy of the obtained eigenvalues is up to 5 digits. It is very similar to calculate the eigenvalue of negative ion H$^{-}$ by setting the nuclear charge $Z=1$ in Eq.(\ref{E20}). It is known that there is only one bound state in H$^{-}$\cite{Pekeris}. Our calculated energy $E=-0.52774$ a.u. agrees well with the accurate energy $E=-0.52775$ a.u. The results demonstrate that our method can be directly used in calculating bound states of three-body systems. However, the calculations of the correlated two-electron continuum states are still challenging.

\begin{table}
  \centering
  \caption{Comparison of the eigenvalues of He S states obtained by ITP method (the first entry) and the values in Ref.\cite{Drake} (the second entry).}\label{T5}

\begin{tabular}{c c c }
\hline
State & Singlet (a.u.)& Triplet (a.u.)\\
\hline
1S&-2.90370&\\
  &-2.90372&\\
2S&-2.145972&-2.17522935\\
&-2.145974&-2.17522937\\
3S&-2.0612712&-2.06868904\\
&-2.0612719&-2.06868906\\
4S&-2.03356&-2.03650\\
&-2.03358&-2.03651\\
\hline
\end{tabular}

\end{table}

\section{Conclusion}\label{V}
In summary, based on the Crank-Nicolson numerical method, we present in this work an extension of the ITP method $t\rightarrow -it$ for directly converging an arbitrary initial vector to desired bound excited states by controlling the time step size. It is different from the usual ITP methods which can only converge directly to the ground state. To obtain converged excited state, the usual ITP methods require filtering out all lower bound states from the initial state. For continuum states with positive energies, we generalize the imaginary propagation method by $t\rightarrow it$, and show convergence of any arbitrary state to desired continuum states. This method is proven to be simple and accurate for small number of iterations $j$ or total time $t=j\Delta t$. The application of this ITP method in the photoionization of diatomic molecules is illustrated. This method can also be used in two-electron systems. We emphasize that this C-N method is a systematic method to study ultrafast dynamics. After obtaining the initial bound state by C-N ITP, then the C-N method can be used to solve TDSE in real time space. After the interactions between intense ultrashort laser pulses and atoms and molecules, the energy and momentum information of the photoelectrons can be extracted from the time-dependent wave functions\cite{Madsen} by projecting on the continuum states obtained by ITP. This new method can be directly extended to complex system\cite{Brosolo2}, confined systems\cite{Ndengu}, Dirac equations\cite{Lorin}, multiparticle states\cite{ZhaL} and also in TDDFT methods\cite{Lopata,Fowe} for complex molecular systems.
\section{ACKNOWLEDGEMENTS}\label{VI}
The author thanks Prof. Andr\'{e} D. Bandrauk very much for helpful discussions.

\end{document}